\documentclass[twocolumn, 	
 showpacs, 			
 preprintnumbers, 		
 aps,  				
 prd,  				
 a4paper, 			
 superscriptaddress, 		
 nofootinbib, 			
 tightenlines, 			
 floats 			
 ]{revtex4}

\usepackage{graphicx}
\usepackage{amssymb,latexsym}
\usepackage[draft=false]{hyperref}

\input epsf

\newcommand{\ba}{\begin{array}}
\newcommand{\ea}{\end{array}} 

\begin{document}

\renewcommand{\topfraction}{0.99}

\title{The simplest curvaton model} 

\author{N.~Bartolo}
\affiliation{Dipartimento di Fisica di Padova ``G.~Galilei'', Via Marzolo 8, 
Padova I-35131, Italy}
\affiliation{INFN, Sezione di Padova, Via Marzolo 8, Padova I-35131, Italy}
\affiliation{Astronomy Centre, University of Sussex, Brighton BN1 9QJ, United 
Kingdom}

\author{Andrew R.~Liddle}
\affiliation{Astronomy Centre, University of Sussex, Brighton BN1 9QJ, United 
Kingdom}

\date{\today} 

\pacs{98.80.Cq, 98.70.Vc \hfill astro-ph/0203076}

\preprint{astro-ph/0203076}

\begin{abstract}
We analyze the simplest possible realization of the curvaton scenario, where a 
nearly scale-invariant spectrum of adiabatic perturbations is generated by 
conversion of an isocurvature perturbation generated during inflation, rather 
than the usual inflationary mechanism. We explicitly evaluate all the 
constraints on the model, under both the assumptions of prompt and delayed 
reheating, and outline the viable parameter space.
\end{abstract}

\maketitle

\section{Introduction}

There has recently been renewed interest in an alternative inflationary 
mechanism for generating an approximately scale-invariant adiabatic density 
perturbation spectrum to the usual one. Rather than immediately generating a 
curvature perturbation via perturbations in the inflaton field \cite{LL}, 
instead the mechanism relies on isocurvature perturbations in another scalar 
field whose energy density is subdominant during inflation. After inflation ends 
this second scalar field comes to contribute significantly to the energy 
density, at which point 
the isocurvature perturbation converts to adiabatic even on superhorizon scales. 
Subsequent complete decay of this second field guarantees purely adiabatic 
perturbations, though variants on this scenario can leave a residual 
isocurvature component too. This mechanism of conversion from isocurvature to 
adiabatic perturbations was
first discussed long ago by Mollerach \cite{Moll}, was briefly mentioned by 
Linde and Mukhanov \cite{LinMuk} in a paper primarily focussing on scenarios for 
nongaussian isocurvature perturbations, and more recently received renewed 
attention in Refs.~\cite{ES,LW,MT}. Amongst these, Lyth and Wands \cite{LW} 
considered the scenario in the broadest context, and named the second scalar 
field the {\em curvaton}.

While Lyth and Wands described the development of the perturbations in 
considerable detail, they sought to keep their discussion as model-independent 
as possible and did not discuss a specific inflationary scenario. In this paper 
we give a specific realization of the curvaton scenario and to evaluate 
all the constraints on model building that need to be satisfied for a successful 
scenario.

\section{The simplest curvaton model}

\label{s:simplest}

While the general curvaton scenario allows perturbation generation featuring 
possible isocurvature components and possible nongaussianity, our aim here is to 
construct a simple curvaton model which creates a nearly scale-invariant 
spectrum of gaussian and purely adiabatic perturbations.
The simplest possible curvaton model features two massive non-interacting scalar 
fields, giving a potential
\begin{equation}
V(\phi,\sigma) = \frac{1}{2} M^2 \phi^2 + \frac{1}{2} m^2 \sigma^2 \,,
\end{equation}
where we indicate the inflaton by $\phi$ and the curvaton by $\sigma$. The 
scenario requires that the curvaton energy density contributes negligibly during 
inflation (in particular the final stages).
Since generically the curvaton 
will end up being the lighter of the two fields, it turns out that the curvaton 
must be close to its minimum, in order to prevent it driving a second period of 
inflation after the $\phi$-driven inflation is complete.
Under these constraints, $\sigma$ will remain constant to a good approximation 
during the later 
stages of inflation, and we indicate this constant value by $\sigma_*$. This 
initial condition is not fixed by the theory, but rather represents an 
additional free parameter to be fixed by observations.\footnote{Under this 
condition 
the oscillation mechanism of Ref.~\cite{BMR} is highly suppressed, and no 
mixture of correlated adiabatic and isocurvature perturbations will
be relevant at the end of inflation.} The curvaton's 
subdominance requires 
$m^2 \sigma_*^2 \ll M^2 \phi^2$ during inflation (where we will have 
$\phi \gtrsim m_{{\rm Pl}}$ where $m_{{\rm Pl}}$ is the Planck mass).

The early stages of inflation set the global mean of $\sigma_*$ in our 
observable Universe and arrange its classical homogeneity, and it then receives 
perturbations via the usual quantum mechanism, with the typical perturbation 
accrued in a Hubble time being $\delta \sigma \simeq H/2\pi$. In order for the 
eventual curvature perturbation to be gaussian, this perturbation must be small 
compared to the mean value of the field, and so we require $\sigma_*^2 \gg 
H^2/4\pi^2$. For these quantum perturbations to be the dominant influence on the 
curvaton, it must be effectively massless during inflation which requires $m^2 
\ll H^2 \simeq 4\pi M^2\phi^2/3m_{{\rm Pl}}^2$.

Once inflation is over and the inflaton energy density converted to radiation, 
the $\sigma$ field will continue to remain constant while its mass is negligible 
compared to the Hubble parameter. Once $m^2 \simeq H^2$, it will begin to 
oscillate about the minimum of its potential, its energy density decaying at an 
average rate of $\rho_\sigma \propto 1/a^3$  (the Universe will have to still be 
radiation dominated by this stage, as otherwise the domination of $\sigma$ will 
initiate a new period of inflation).

The final stage is the decay of the curvaton, which in this paper we will assume 
is a complete decay into conventional matter which thermalizes with the existing 
radiation. The decay occurs on a timescale 
$\Gamma_\sigma$, which is a further free parameter of the scenario. The most 
conservative constraint on the decay rate is that conventional radiation 
domination had better be in place by nucleosynthesis, though baryogenesis 
scenarios are likely to require a much earlier decay. Decay will happen when 
$\Gamma_\sigma \simeq H$, so this requires $\Gamma_\sigma > H_{{\rm nuc}} \simeq 
10^{-40} 
m_{{\rm Pl}}$.
The time until decay sets the magnitude of the curvature perturbation generated, 
because it determines what fraction of the mean energy density comes from the 
curvaton when it decays. We require to match the COBE normalization; this sets a 
minimum requirement on the size of the curvaton fluctuations because they must 
be large enough to generate the required perturbations in the limit where the 
curvaton field is completely dominant when it decays. The inflaton will also 
generate a curvature perturbation at some level, and while a mixed perturbation 
scenario is permitted we will only consider here the case where the 
inflaton-generated perturbation is negligible, which requires $M \ll 10^{-6} 
m_{{\rm Pl}}$.

\section{Model constraints}

\subsection{The case of prompt reheating}

In this subsection, we shall assume that after inflation ends reheating occurs 
promptly, with the inflaton decaying into radiation. In that case, no further 
parameters are necessary to specify the scenario; we have four parameters which 
are the two masses $m$ and $M$, the curvaton value during inflation $\sigma_*$, 
and the curvaton decay constant $\Gamma_\sigma$. 

In a quadratic potential, inflation ends by violation of slow-roll at 
$\phi_{{\rm end}} \simeq m_{{\rm Pl}}/\sqrt{4\pi}$, corresponding to a Hubble 
parameter $H_{{\rm end}}^2 = M^2/3$. During the subsequent radiation-dominated 
era $H^2 = H_{{\rm end}}^2 a_{{\rm end}}^4/a^4$ where $a$ is the scale factor.

The next event to take place is for the curvaton to become effectively massive, 
$m^2 = H^2$. This happens when 
\begin{equation}
\left( \frac{a_{{\rm mass}}}{a_{{\rm end}}} \right)^4 = \frac{M^2}{3m^2} \,.
\end{equation}
In order to prevent a period of curvaton-driven inflation, the Universe must 
still be radiation dominated at that point, which implies a significant 
constraint
\begin{equation} \label{nci}
\sigma_*^2 \ll \frac{3}{4\pi} \, m_{{\rm Pl}}^2\,.
\end{equation}
This is a substantial restriction amongst all the possible values that $\sigma$ 
might have taken (most of which would result in a long epoch of $\sigma$-driven 
inflation after the $\phi$ field has reached its minimum).

The most important constraint on the parameters is the requirement of 
reproducing the observed perturbation amplitude. Denoting the ratio of the 
curvaton energy density to that of radiation by $r \equiv \rho_\sigma/\rho_{{\rm 
rad}}$, in the limit where $r < 1$ (i.e.~the curvaton decays during radiation 
domination), Lyth and Wands \cite{LW} 
demonstrated that the spectrum of the Bardeen parameter ${\cal P}_\zeta$, whose 
observed value is about $2 \times 10^{-9}$, is given by 
\begin{equation} \label{S}
{\cal P}_\zeta \simeq \frac{r_{{\rm decay}}^2}{16} \frac{H_*^2}{\pi^2
	\sigma_*^2} \,,
\end{equation}
where
\begin{equation}
H_*^2 \simeq \frac{100}{3} \, M^2  
\end{equation}
is the Hubble parameter when observable perturbations were generated, around 50 
$e$-foldings before the end of inflation, and
\begin{equation}
r = \frac{\rho_\sigma^{{\rm end}}}{\rho_{{\rm rad}}^{{\rm end}}} \,
	\left( \frac{a_{{\rm mass}}}{a_{{\rm end}}} \right)^4 \,
	\frac{a}{a_{{\rm mass}}} = \frac{4\pi}{3} \frac{\sigma_*^2}{m_{{\rm
	Pl}}^2} \, \frac{a}{a_{{\rm mass}}} \,.
\end{equation}
Continuing to presume that the Universe is still radiation dominated at decay, 
this will be at $\Gamma_\sigma^2 = H^2 = H^2_{{\rm end}} a_{{\rm end}}^4/a^4$, 
giving 
\begin{equation} \label{r}
r_{{\rm decay}} = \frac{4\pi}{3} \, \frac{\sigma_*^2}{m_{{\rm Pl}}^2} \,
	\sqrt{\frac{m}{\Gamma_\sigma}} \,,
\end{equation}
and hence
\begin{equation}
\label{pzeta1}
{\cal P}_\zeta \simeq 4 \, \frac{mM^2 \sigma_*^2}{\Gamma_\sigma m_{{\rm Pl}}^4} 
	\quad \mbox{(for $r_{{\rm decay}} < 1$)} \,.
\end{equation}
We find that achieving the correct perturbation amplitude, in combination with 
other constraints, excludes all the regions of parameter space where the 
curvaton decays while still effectively massless.

In the opposite regime, where $r_{{\rm decay}}$ exceeds one and so decay occurs 
after curvaton domination, this formula no longer holds and instead the 
perturbation produced becomes independent of $\Gamma_\sigma$, being \cite{LW}
\begin{equation} \label{P2}
\label{pzeta2}
{\cal P}_\zeta \simeq \frac{1}{9} \, \frac{H_*^2}{\pi^2 \sigma_*^2} \simeq
	\frac{M^2}{4 \sigma_*^2} \quad \mbox{(for $r_{{\rm decay}} > 1$)}\,.
\end{equation}
The two expressions agree at the transition $r_{{\rm decay}} \sim 1$. Note that 
if this last expression is normalized as quoted above, the gaussianity condition 
$\sigma_*^2 \gg H_*^2/4\pi^2 \simeq M^2$ is automatically satisfied.

\begin{figure*}
\includegraphics[width=8cm]{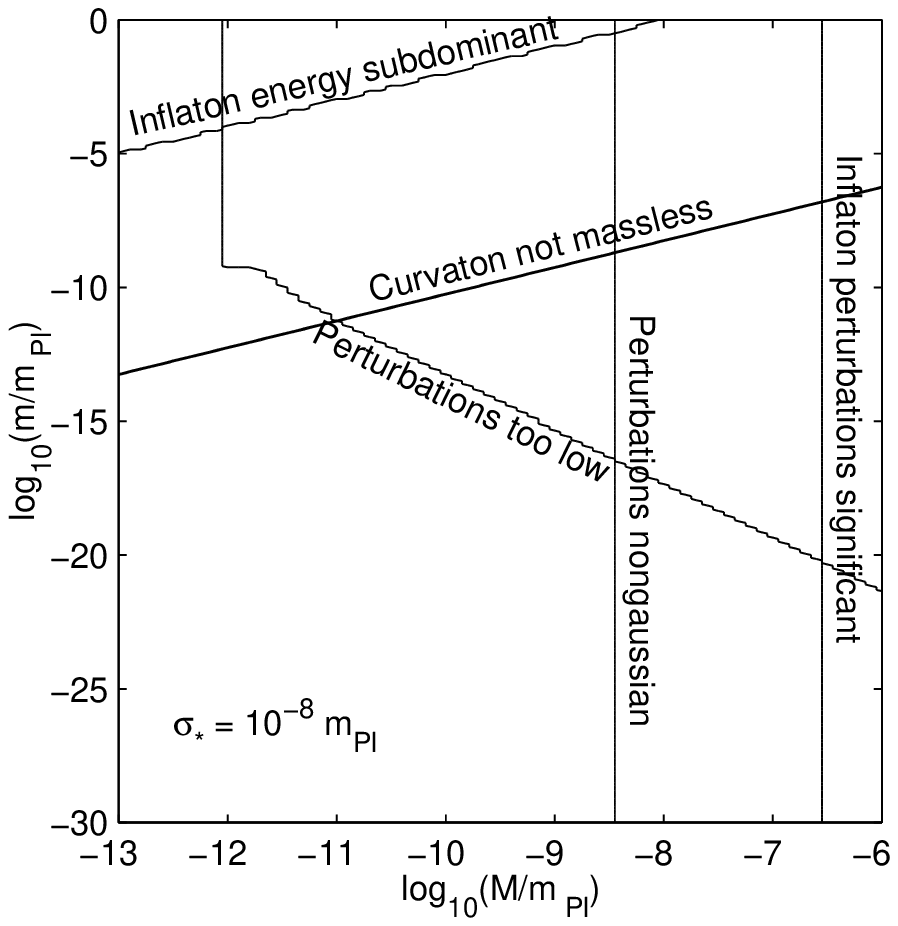}
\hspace*{0.4cm} \includegraphics[width=8cm]{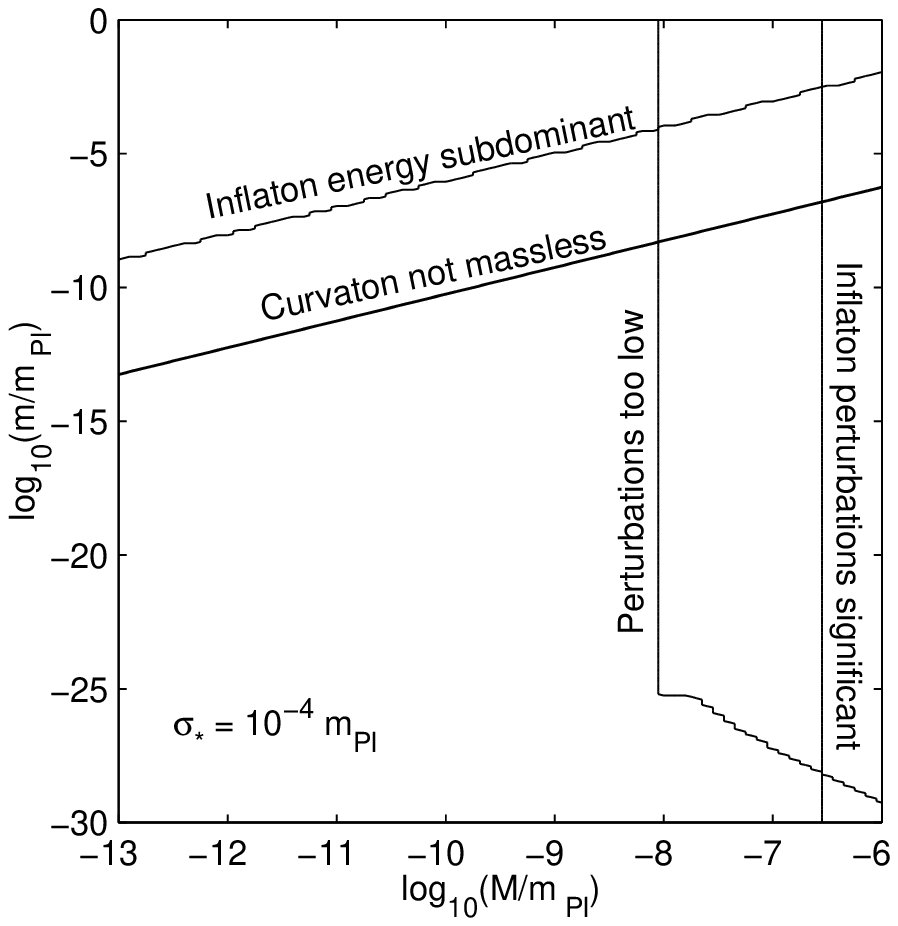}
\caption{\label{fig:prompt1} Allowed regions of parameter space for 
$\sigma_*=10^{-8} m_{{\rm Pl}}$ (left panel) and $\sigma_*=10^{-4}m_{{\rm Pl}}$ 
(right panel). In all 
cases, the constraint lines are identified with a label on the outside of the 
allowed region.}
\end{figure*}

With four parameters to vary and only one equality, Eq.~(\ref{pzeta1}) or
(\ref{pzeta2}), imposed upon them, we expect quite a bit of freedom in choosing
suitable parameters.  However the set of inequalities the parameters must
satisfy is a large one, and it turns out that viable parameter space is quite
restricted.  For definiteness, we set thresholds that the inflaton perturbation
must be no more than ten percent of the curvaton perturbation (i.e.~$M < 3
\times 10^{-7} m_{{\rm Pl}}$) and that the gaussianity condition $\sigma_*^2 \gg
H_*^2/4\pi^2$ on observable scales be satisfied by an order of magnitude.

We proceed by considering triplets of values $(m,M,\sigma_*)$. For such a 
triplet, we fix $\Gamma_\sigma$ using the power spectrum normalization. In some 
areas of parameter space this cannot be achieved, either because the curvaton 
perturbations are too small even if the curvaton energy density becomes 
dominant, or because one would violate $\Gamma_\sigma \gtrsim 
10^{-40} m_{{\rm Pl}}$ as required by nucleosynthesis. Otherwise, we then test 
whether the many other requirements to build a successful model are achieved, 
namely we must guarantee that the inflaton dominates during inflation, that the 
curvature perturbation from the inflaton is negligible, that the curvaton is 
effectively massless 
during inflation and that the curvature perturbations resulting from the 
curvaton are gaussian, and that the curvaton energy density is still subdominant 
when it begins oscillating. This set of constraints slices off regions of the 
parameter space, leaving the region in which viable models can be constructed.

Fig.~\ref{fig:prompt1} shows the allowed regions for two choices of $\sigma_*$, 
with the main constraints plotted. The perturbation amplitude constraint has two 
branches; the vertical part indicates that the models cannot reach the required 
perturbation amplitude even once the curvaton becomes fully 
dominant,\footnote{On the vertical line itself lie models where the curvaton can 
dominate at decay; these are best analyzed separately which we do in the 
following paragraphs.} while the lower part of the curve indicates that the 
perturbations have not grown sufficiently by nucleosynthesis. For low 
$\sigma_*$, the nongaussianity constraint sweeps leftwards across the allowed 
region and there are no viable models once $\sigma_* \lesssim 10^{-10} m_{{\rm 
Pl}}$. For 
high $\sigma_*$ it becomes impossible to generate sufficient perturbations, 
cutting off parameter space above $\sigma_* \simeq 3 \times 10^{-3} m_{{\rm 
Pl}}$. There is 
however a significant parameter space of viable models satisfying all our 
requirements.

For the special case of the curvaton decaying when its energy density is 
dominating over radiation, an analytical analysis of the parameter space is 
possible. In this case $\Gamma_\sigma$ no longer enters into the relevant 
expression for ${\cal P}_\zeta$, Eq.~(\ref{P2}). Thus, using the power spectrum 
normalization to fix the value of $\sigma_{*}$ in terms of $M$ as 
$\sigma_{*}^2=1.2 \times 10^8\, M^2$, the relevant constraints reduce to
\begin{equation} 
\label{c1} 
M<3 \times 10^{-7} \, m_{{\rm Pl}}\, , \quad m^2 \ll \frac{M^2}{3}
\end{equation} 
and 
\begin{equation} \label{c2}
H_{{\rm nucl}} \simeq 10^{-40} m_{{\rm Pl}} < \Gamma_\sigma <
	\left(\frac{4\pi}{3}\right)^2 \, m\, \left( \frac{\sigma_{*}}{m_{{\rm 
Pl}}}
	\right)^4\, .
\end{equation}
The first two conditions require that the inflaton-generated perturbations be 
subdominant and that the curvaton be massless during inflation respectively, and 
are the same as before. If they are satisfied, the remaining constraints  
automatically follow. 
The condition on $\Gamma_\sigma$ derives from requiring  $r_{{\rm decay}}>1$, 
where in this case
\begin{equation}
r_{{\rm decay}}=\left( \frac{4\pi}{3} \right)^{4/3}
\left( \frac{\sigma_*}{m_{{\rm Pl}}} \right)^{8/3}
\left( \frac{m}{\Gamma_\sigma} \right)^{2/3}\, ,
\end{equation} 
as a consequence of the $\sigma$ decay after curvaton domination.

\begin{figure}[t]
\includegraphics[width=8cm]{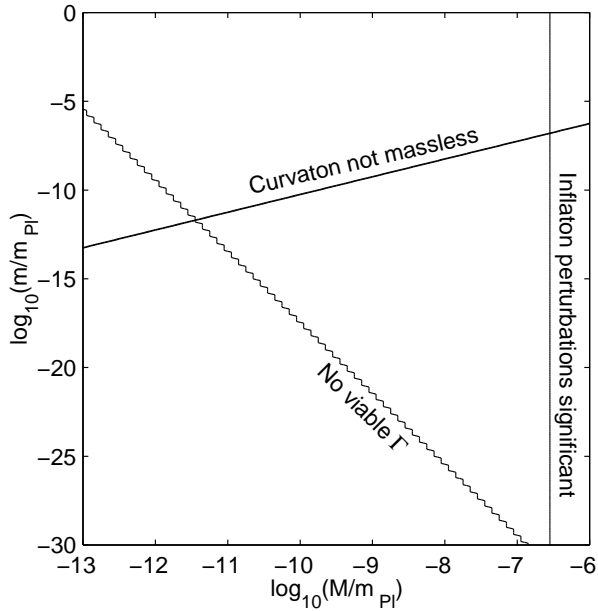}\\
\caption{\label{fig:prompt2} The parameter space in the case of the curvaton 
dominating before decay. Models within the triangular region are viable; they 
have $\sigma_*^2=1.2 \times 10^8\, M^2$, and $\Gamma_\sigma$ may take on any 
value in the range given by Eq.~(\ref{c2}).}
\end{figure}

The formulae Eqs.~(\ref{c1}) and (\ref{c2}) give an allowed region in the
$m$--$M$ plane shown in Fig.~\ref{fig:prompt2}.  At each point within the
allowed region there is a range of permitted $\Gamma_\sigma$ indicated by
Eq.~(\ref{c2}) which gives the correct density perturbation normalization.
These regions correspond to the vertical part of the perturbation amplitude
constraint in Fig.~\ref{fig:prompt1} for the corresponding $\sigma_*$.

\subsection{The case of prolonged reheating}

Reheating is not expected to be instantaneous, and in this subsection we 
generalize the previous results to allow for a delay before reheating is 
complete. However there are no qualitative differences to the scenario and so we 
will keep the discussion fairly brief.

In the case of prolonged reheating, after inflation there is a significant 
period during which the inflaton oscillates coherently at the bottom of its 
potential. Its ultimate decay products will be considered much lighter than 
$\phi$ itself thus constituting the radiation, and we assume that $\phi$ decays 
into radiation with a rate $\Gamma_{\phi}$. We make the assumption that there 
are no significant decays of the inflaton into the curvaton 
field.\footnote{Were there such decays, they would typically
generate an extra, nearly homogeneous, component of the curvaton energy density, 
which might affect the ability of the curvaton to generate sufficiently large 
perturbations.} 
The new parameter $\Gamma_{\phi}$ enters to modify the previous constraints. 

To prevent a new period of inflation driven by the curvaton, its energy density 
must still be subdominant during reheating. Following some standard 
approximations \cite{KT}, we can say that the inflaton $\phi$ starts oscillating 
at the end of inflation (when $H^2_{{\rm end}} = M^2/3$) and
it behaves as $\rho_{\phi} \propto a^{-3}$ until $H \approx \Gamma_{\phi}$, 
when it decreases exponentially ($\rho_{\phi} \propto e^{-\Gamma_{\phi}t}$) 
producing most of the radiation. When $H \approx \Gamma_{\phi}$ reheating is 
completed. Before that point some radiation will be produced, but it is 
subdominant and $\rho_{{\rm rad}}$ scales as $a^{-3/2}$. After reheating the 
universe will be radiation dominated. 

Concerning the curvaton we can consider two situations, one where 
$\sigma$ starts oscillating \emph{after} reheating, and the other where it 
begins oscillating \emph{during}  reheating. Just for illustrative purposes we 
assume that $\sigma$ finally decays with a rate $\Gamma_{\sigma}$ after 
reheating when it is still subdominant with respect to the produced radiation. 
Then $P_{\zeta}$ is given by Eq.~(\ref{S}) with $r_{\text{decay}}<1$. The 
constraints defining the curvaton model during inflation, as described in 
Sec.~\ref{s:simplest}, still hold as does the requirement $\Gamma_{\sigma} > 
H_{{\rm nucl}} \simeq 10^{-40} m_{{\rm Pl}}$. When $\sigma$ starts oscillating 
it should be a subdominant component in order to prevent a period of 
curvaton-driven inflation, and it can be checked that this constraint remains 
the same as Eq.~(\ref{nci}).

In the case where the oscillations of the curvaton start after reheating 
$\Gamma_{\phi} > m$ and 
\begin{equation}
\left( \frac{a_{{\rm reh}}}{a_{{\rm mass}}} \right)^4 = 
\frac{m^2}{\Gamma_{\phi}^2} \,,
\end{equation}
where $a_{{\rm reh}}$ is the scale factor at the end of reheating.

On the other hand the produced radiation has an energy density $\rho_{{\rm 
rad}}^{{\rm reh}}=3m_{{\rm Pl}}^2\Gamma_{\phi}^2/8\pi$. As a consequence the 
expression for $r_{{\rm decay}}$ is exactly the same of Eq.~(\ref{r}), and it is 
independent of $\Gamma_{\phi}$
\begin{equation} 
\label{r1}
r_{{\rm decay}} = \frac{4\pi}{3} \,\sigma_*^2 \frac{m^2}{m_{{\rm Pl}}^2} \,
	\sqrt{\frac{1}{m^3 \Gamma_{\sigma}}} \,.
\end{equation} 
Note, however, that if $\Gamma_{\phi} > m$ and the constraint in Eq.~(\ref{nci}) 
is satisfied, then 
\begin{equation} 
\label{sc}
\frac{4\pi}{3} \,\sigma_*^2 \frac{m^2}{m_{{\rm Pl}}^2} \ll \Gamma_{\phi}^2 
\,.
\end{equation}
This amounts to saying that throughout reheating the energy density of the 
curvaton has to be much smaller than the radiation energy density at the end of 
reheating. The strong constraint in Eq.~(\ref{sc}) implies that the ratio $r$ at 
the moment of the curvaton decay can indeed be much smaller than in the case of 
prompt reheating. In fact this case is recovered when 
$\Gamma^2_{\phi}=H^2_{{\rm end}} = M^2/3$, but for a prolonged period of 
reheating $\Gamma^2_{\phi} < H^2_{{\rm end}}$. Thus a similar analysis of the 
parameter space could be done as in the previous subsection, taking into account 
now that $m^2< \Gamma^2_{\phi} < M^2/3$. 

The curvaton begins oscillating during reheating when
\begin{equation}
\left ( \frac{a_{{\rm mass}}}{a_{{\rm end}}} \right)^3=\frac{M^2}{3m^2} \, .
\end{equation}
In the approximation $\rho_{{\rm rad}}^{{\rm reh}} \simeq \rho_{\phi}^{{\rm 
end}} (a_{{\rm 
end}}/a_{{\rm reh}})^3$, one finds
\begin{equation} \label{r2}
r_{{\rm decay}} = \frac{4\pi}{3} \frac{\sigma_*^2}{m_{{\rm Pl}}^2}
	\sqrt{\frac{\Gamma_{\phi}}{\Gamma_{\sigma}}} \,.
\end{equation}
Since $\Gamma_{\phi}<m$, a comparison with Eq.~(\ref{r1}) shows that $r_{{\rm 
decay}}$ can be even smaller than in the case when $\sigma$ starts oscillating 
after reheating: once $\sigma_{*}, \, \Gamma_{\sigma}\, \rm{and}\,  
\Gamma_{\phi}$ are fixed, the curvaton energy density starts decreasing as 
$a^{-3}$ at an earlier time.
        
\section{Conclusions}

The curvaton model is an interesting new proposal for generating the 
approximately scale-invariant curvature perturbations which presently give the 
best match to observational data. While nongaussianity would not necessarily be 
an observational disaster, we have chosen to restrict our attentions to gaussian 
models.
We have constructed the simplest possible 
realization of this idea, and demonstrated how various requirements close off 
regions of parameter space while leaving a substantial area of viable models. 
Even this simplest model, with prompt reheating, features four parameters, and 
so the predicted perturbations in different regions of parameter space are 
highly degenerate.

While three of these parameters ($m$, $M$ and $\Gamma_\phi$) are parameters of 
the underlying theory, the fourth, $\sigma_*$, refers to the initial conditions 
for our patch of the Universe. The required values of $\sigma_*$ for a 
successful curvaton model have magnitude less than about $10^{-3} \, m_{{\rm 
Pl}}$, which represents only a small region of the plausible values for 
$\sigma_*$ that might exist during inflation, as its potential has too low a 
magnitude to influence the dynamics. Nevertheless, in a typical chaotic 
inflation scenario one expects regions which do satisfy this criterion by 
chance. If one wishes, there are also opportunities to introduce anthropic 
principle considerations; for fixed values of $m$ and $M$, regions with large 
$\sigma_*$ would typically lead to the inflaton evolving to the bottom of its 
potential followed by a period of slow-roll inflation driven by the curvaton, 
which given the small value of $m$ would generate a much lower level of 
curvature perturbations than in the curvaton domain and hence not give rise to 
structure in the Universe.

As far as the density perturbations are concerned, the predictions of these 
models are indistinguishable from slow-roll inflation models arranged to give 
the same spectral index. However, the curvaton model has the feature that the 
gravitational wave amplitude is predicted to be low, because the Hubble rate 
during inflation is less than in an equivalent slow-roll model. In particular, 
the scalar and tensor perturbations typically will not obey the usual 
consistency relation (see e.g.~Ref.~\cite{LL}). In fact, in a natural curvaton 
model where $\sigma$ is extremely light ($m^2 \ll H^2$) and subdominant with 
respect to the inflaton, the scalar and tensor indices are predicted to be the 
same: $n_{s}-1=n_{t}+2m^2/3H^2_{*} \simeq n_{t}$ \cite{LW}. Equal spectral 
indices is a prediction of the power-law class of conventional inflation models, 
but they predict a high amplitude of gravitational waves, contrary to the 
curvaton model. Unfortunately however the low amplitude of tensors in the 
curvaton model makes it difficult to detect the tensors at all, and one cannot 
expect a useful measure of their spectral index. 

While one can 
certainly design slow-roll inflation models with any spectral index for the 
density perturbations and 
negligible gravitational waves, so that the curvaton model predictions cannot be 
viewed as distinct, a detection of a significant amplitude of gravitational 
waves would be sufficient to rule out these curvaton models. It is however 
interesting to note that present observations of large-scale 
structure and cosmic microwave background anisotropies tend to prefer a slight 
red tilt and a low contribution of tensor modes \cite{WTZ}, as predicted by 
models of the type we have discussed.

\section*{Acknowledgments}

N.B.~was partially supported by a Marie Curie Fellowship
of the European Community programme HUMAN POTENTIAL under contract  
HPMT-CT-2000-00096, and A.R.L.~in part by the Leverhulme Trust. 
We thank David Lyth and David Wands for discussions.

 

\end{document}